\documentclass[12pt,a4paper]{article}
\usepackage{a4wide}
\usepackage{latexsym}
\usepackage{epsf}
\usepackage{amssymb}

\usepackage{amsmath}
\usepackage{slashed,epsfig}

\makeatletter
\@addtoreset{equation}{section}
\makeatother

\newsavebox{\uuunit}
\sbox{\uuunit}
    {\setlength{\unitlength}{0.825em}
     \begin{picture}(0.6,0.7)
        \thinlines
        \put(0,0){\line(1,0){0.5}}
        \put(0.15,0){\line(0,1){0.7}}
        \put(0.35,0){\line(0,1){0.8}}
       \multiput(0.3,0.8)(-0.04,-0.02){12}{\rule{0.5pt}{0.5pt}}
     \end {picture}}
\newcommand {\unity}{\mathord{\!\usebox{\uuunit}}}

\begin{document}

\begin{flushright}
\small
CERN-TH/2003-035\\
IFT-UAM/CSIC-03-03\\
UG-03-01\\
{\bf hep-th/0303113}\\
\date \\
\normalsize
\end{flushright}

\begin{center}


\vspace{.7cm}

  {\LARGE {\bf Domain Walls of $D=8$ Gauged Supergravities and their 
$D=11$ Origin}} \\

\vspace{1.2cm}

{\large Natxo Alonso-Alberca}${}^{\spadesuit,\heartsuit}$
,
{\large Eric Bergshoeff}${}^{\diamondsuit}$
,
{\large Ulf Gran}${}^{\diamondsuit}$
, \\
{\large Rom\'an Linares}${}^{\diamondsuit}$
,
{\large Tom\'as Ort\'{\i}n}${}^{\spadesuit,\clubsuit}$
{\large and}
{\large Diederik Roest}${}^{\diamondsuit}$
\vskip 1truecm

\small
${}^{\spadesuit}$\ {\it Instituto de F\'{\i}sica Te\'orica, C-XVI,
Universidad Aut\'onoma de Madrid \\
E-28049-Madrid, Spain}
\vskip 0.2cm
${}^{\heartsuit}$\ {\it Departamento de F\'{\i}sica Te{\'o}rica, C-XI,
Universidad Aut\'onoma de Madrid\\
Cantoblanco, E-28049 Madrid, Spain}
\vskip 0.2cm
${}^{\diamondsuit}$\ {\it Centre for Theoretical Physics, University of Groningen,\\
   Nijenborgh 4, 9747 AG Groningen, The Netherlands.}
\vskip 0.2cm
${}^{\clubsuit}$\ {\it Theory Division, C.E.R.N., CH-1211, Geneva 23, Switzerland}

\vspace{.7cm}


{\bf Abstract}

\end{center}

\begin{quotation}

\small

Performing a Scherk-Schwarz dimensional reduction of $D=11$ supergravity on a 
three-dimensional group manifold we construct five $D=8$ gauged maximal
supergravities whose gauge groups are the three-dimensional (non-)compact 
subgroups of $SL(3,\mathbb{R})$. These cases include the Salam-Sezgin $SO(3)$ 
gauged supergravity. We construct the most general half-supersymmetric domain 
wall solutions to these five gauged supergravities. The generic form is a 
triple domain wall solution whose truncations lead to double and single domain 
wall solutions. We find that one of the single domain wall solutions has zero 
potential but nonzero superpotential.

Upon uplifting to 11 dimensions each domain wall becomes a purely 
gravitational 1/2 BPS solution. The corresponding metric has a $7+4$ split 
with a Minkowski 7-metric and a 4-metric that corresponds to a gravitational 
instanton. These instantons generalize the $SO(3)$ metric of Belinsky, 
Gibbons, Page and Pope (which includes the Eguchi-Hanson metric) to the other 
Bianchi types of class A.

\end{quotation}

\newpage

\pagestyle{plain}

\section{Introduction}

Gauged supergravities have become the focus of recent research due to a 
variety of reasons. Most applications are related to the fact that gauged 
supergravities contain a nonzero potential for the scalar fields. This 
potential, which behaves as an effective cosmological constant, allows for 
interesting vacuum solutions such as de Sitter and anti-de Sitter spacetimes 
or (half-supersymmetric) domain wall solutions. The anti-de Sitter vacua are 
important in the context of the AdS/CFT correspondence \cite{Maldacena:1998re} 
while the possibility of a de Sitter vacuum is of interest due to recent 
astronomical observations \cite{AdS}. This has triggered a search for de 
Sitter vacuum solutions in string theory (see, e.g., 
\cite{Fre:2002pd,deRoo:2002jf}). Domain wall solutions have applications to 
the DW/QFT correspondence \cite{Itzhaki:1998dd,Boonstra:1998mp}, the 
braneworld scenario \cite{Randall:1999ee,Randall:1999vf} and cosmological 
models (see, e.g., \cite{Kallosh:2001gr,Townsend:2001ea}).

We are particularly interested in domain wall solutions in $D=5$ dimensions in 
connection with the search for a supersymmetric braneworld scenario. Finding 
the most general 1/2 BPS domain wall solution in $D=5$ dimensions is a 
difficult task due the fact that the scalar potential is a complicated 
function of the many scalars that are present in the theory. To learn more 
about the $D=5$ situation, it is instructive to study the simpler case of 
general 1/2 BPS domain wall solutions of maximal supergravities in higher 
dimensions $D \le 11$. The first nontrivial example is $D=9$ and this case was 
already discussed in 
\cite{Cowdall:2000,Bergshoeff:2002mb,Nishino:2002zi,Bergshoeff:2002nv}. The 
aim of this paper is to study the situation in $D=8$ dimensions.

The standard $D=8$ gauged maximal supergravity is the
$SO(3)$-gauged theory of Salam and Sezgin \cite{Salam:1985ft}. They
constructed this theory by applying what we will call a 
Scherk-Schwarz 2 (SS2) reduction procedure \cite{Scherk:1979zr} to $D=11$ 
supergravity. The SS2 procedure corresponds to a reduction on a group manifold
where one uses a {\it symmetry of the compactification manifold}
to give a 
specific dependence of the $D=11$ fields on the compactification 
coordinates. This dependence is such that, although the $D=11$ fields depend 
on them, the resulting $D=8$ action does not. 
In contrast, there is also a so-called
SS1 procedure \cite{Scherk:1979ta} where the higher-dimensional fields 
acquire a dependence on the compactification coordinates by using a 
{\it global, internal symmetry} of the higher-dimensional theory, such as the 
$SL(2,\mathbb{R})$ symmetry of Type IIB supergravity.

In this paper we repeat and generalize the analysis of \cite{Salam:1985ft} to 
a group manifold corresponding to an arbitrary Lie algebra, instead of $SO(3)$ 
only. We will show that the standard Bianchi classification of 
three-dimensional Lie algebras, see e.g.~\cite{Wald:1984rg}, leads to five 
cases, including the $S^3$ group manifold used in \cite{Salam:1985ft}. 
Performing a SS2 reduction of $D=11$ supergravity with respect to these five 
distinct group manifolds we construct five gauged maximal 
supergravities where the gauge groups are the three-dimensional (non-)compact 
subgroups of $SL(3,\mathbb{R})$. These cases include the Salam-Sezgin $SO(3)$ 
gauged supergravity. The other four cases, which involve the gauge groups 
$SO(2,1), ISO(2), ISO(1,1)$ and the Heisenberg group, can be obtained by 
analytic continuation and/or generalized 
In\"on\"u-Wigner contractions of the Salam-Sezgin theory.

We point out that the Bianchi classification allows for five more cases where 
the gauge groups are three-dimensional non-compact subgroups of 
$GL(3,\mathbb{R}) = SL(3,\mathbb{R}) \otimes SO(1,1)$. Due to the extra 
$SO(1,1)$-factor, the group manifold isometries are only symmetries of the 
equations of motion but not of an action. Therefore the SS2 procedure leads in 
these cases to gauged supergravities whose equations of motion cannot be 
integrated to an action. Two of these theories contain a free mass parameter. 
In the limit that this mass parameter goes to zero one recovers two of the 
gauged supergravities that do have an action. To distinguish between 
supergravities having an action or having no action we will denote the ones 
with an action as class A supergravities and the ones without an action as 
class B supergravities, in accordance with the Bianchi classification.

We will present a 1/2 BPS triple domain wall solution that solves the 
equations of motion corresponding to each of the different class A $D=8$ 
gauged supergravities. The truncation of this triple domain wall solution to a 
single domain wall solution reproduces results that partly are already 
available in the literature. In particular, we find a new domain wall solution 
with zero potential but nonzero superpotential. We discuss the uplifting of 
the triple domain wall solution to $D=11$ dimensions and show that, after 
uplifting, it becomes a purely gravitational 1/2 BPS solution. In each case 
the $D=11$ metric has a $7+4$ split with a Minkowski 7-metric and a 4-metric 
that can be identified with a $D=4$ gravitational instanton 
\cite{Hernandez:2002fb}. These instantons generalize the metric of Belinsky, 
Gibbons, Page and Pope (which includes the Eguchi-Hanson metric) to the other 
(class A) Bianchi types.

The organization of this paper is as follows: in Section 2 we perform the SS2 
reduction of $D=11$ supergravity leading to the two classes of $D=8$ gauged 
supergravities mentioned above. In Section 3 we give the action and 
transformation rules of the five different $D=8$ class A gauged 
supergravities. The triple domain wall solution is presented in Section 4. 
Subsequently, in Section 5 the triple domain wall solution is uplifted to 
$D=11$ dimensions and linked to known solutions in the literature. 
Finally, in the Conclusions we discuss several extensions 
and open issues. This paper contains four Appendices. Our conventions are 
given in Appendix~\ref{sec-conventions}. In Appendix B we give a few details 
of $D=11$ supergravity. Appendix~\ref{full8daction} contains the details of 
the (bosonic) action and supersymmetry rules of the class A $D=8$ gauged 
supergravities. Finally, in Appendix D we collect some basic material on the 
classification of three-dimensional Lie algebras.

\section{Reduction on a Group Manifold}

In this Section we perform the reduction of $D=11$ supergravity over a 
three-dimensional group manifold to $D=8$ dimensions. The group manifold reduction procedure 
generally gives rise to gauged supergravities, where the structure constants 
of the gauge group $\cal G$ are provided by the group manifold.

In the case at hand the group manifolds are three-dimensional. The prime 
example is the reduction over the three-sphere $S^3$, which gives rise to the 
Salam-Sezgin $SO(3)$ gauged supergravity \cite{Salam:1985ft}. By choosing 
other structure constants, corresponding to other three-dimensional Lie 
algebras, one can choose other group manifolds, some of which give rise to 
non-compact gaugings. In this section we describe the $D=11$ supergravity 
theory and the reduction Ansatz that leads to the $D=8$ gauged supergravity 
theories. We discuss the classification of the different $D=3$ Lie algebras in 
appendix~\ref{Bianchi} and return to the issue of different $D=8$ 
supergravities in Section~\ref{8dsugras}.

The fields of $N=1,D=11$ supergravity \cite{Cremmer:1978km} are the
Elfbein, a three-form potential and a 32-component Majorana
gravitino{}\footnote{Our conventions are given in
  appendix~\ref{sec-conventions}.}:
\begin{equation}
\rm{11D:~~~}
\left\{\hat{e}_{\hat{\mu}}{}^{\hat{a}},
\hat{C}_{\hat{\mu}\hat{\nu}\hat{\rho}},
\hat{\psi}_{\hat{\mu}}
\right\}\, .
\end{equation}
The bosonic part of the action and the supersymmetry up to bilinear fermions
are given in appendix \ref{11Dsugra}.

To perform the dimensional reduction it is convenient to make an $8+3$ split of the
11-dimensional space-time: $x^{\hat{\mu}} = (x^\mu, z^m)$ with
$\mu=(0,1,\ldots7)$ and $m=(1,2,3)$.  We use the convention that
space-time indices are $\hat{\mu} = (\mu, m)$ while the tangent
indices are $\hat{a} = (a, i)$.  Using a particular Lorentz frame
the reduction Ansatz for the 11-dimensional fields is
\begin{equation}
  \hat{e}_{\hat{\mu}}{}^{\hat{a}}   = 
\left(
\begin{array}{cr}
e^{-\frac{1}{6}\varphi} e_{\mu}{}^{a} & 
e^{\frac{1}{3}\varphi} L_{m}{}^{i}A^{m}{}_{\mu} \\
&\\
0             & 
e^{\frac{1}{3}\varphi}L_{n}{}^{i}\,U^{n}{}_{m}   \\
\end{array}
\right) \, 
\end{equation}
and 
\begin{equation}
\hat{C}_{abc} =  e^{\frac{1}{2}\varphi}\, C_{abc}\, ,
\hspace{.3cm}
\hat{C}_{abi} =  L_{i}{}^{m}B_{m\, ab}\, ,
\hspace{.3cm}
\hat{C}_{aij} = 
 e^{-\frac{1}{2}\varphi}\, \epsilon_{mnp} L_{i}{}^{m} L_{j}{}^{n}\, V_a{}^p \, ,
\hspace{.3cm} \hat{C}_{ijk}= e^{-\varphi}\epsilon_{ijk} \ell \,
\end{equation}
for the bosonic fields and
\begin{equation}
  \begin{array}{rcl}
{\hat \psi}_{\hat a} & = e^{\varphi/12}\left( \psi_{a} - \frac{1}{6} 
\Gamma_{a} \Gamma^{i} \lambda_i \right ) \,, \qquad {\hat\psi}_{i} = 
e^{\varphi/12} \lambda_i \,, \qquad \hat \epsilon = e^{-\varphi/12} \epsilon 
\,
  \end{array}
\end{equation}
for the fermions.  Thus the full 8-dimensional field content consists
of the following 128 + 128 field components (omitting spacetime
indices on the potentials):
\begin{align}
{\rm{8D:~~~}} 
 \{ e_\mu{}^a, L_m{}^i, \varphi, 
  \ell, A^m, V^m, B_m, C, \psi_\mu, \lambda_i \} \,.
\end{align}
All these $D=8$ fields are taken to be independent of $z^m$. We will now 
describe the quantities appearing in this reduction Ansatz.

The matrix $L_m{}^i$ describes the five-dimensional $SL(3,\mathbb{R}) / SO(3)$
scalar coset of the internal space. It transforms under a global $SL(3,\mathbb{R})$ acting from 
the left and a local $SO(3)$ symmetry acting from the right.  We take the
following explicit representative, thus gauge fixing the local $SO(3)$ symmetry:
\begin{align}
  L_m{}^i = \left(
  \begin{array}{ccc}
    e^{-\sigma/\sqrt{3}} & 
    e^{-\phi/2+\sigma/2\sqrt{3}} \chi_1 & e^{\phi/2+\sigma/2\sqrt{3}} \chi_2 \\
    0 & e^{-\phi/2+\sigma/2\sqrt{3}} & e^{\phi/2+\sigma/2\sqrt{3}} \chi_3 \\
    0 & 0 & e^{\phi/2+\sigma/2\sqrt{3}}
\end{array}
\right) \,,
\label{Lscalar}
\end{align}
which contains two dilatons, $\phi$ and $\sigma$, and three 
axions{}\footnote{We call the scalars $\ell, \chi_1, \chi_2$ and $\chi_3$ 
axions and the scalars $\varphi, \phi$ and $\sigma$ dilatons since (in the 
ungauged case) the axions only occur with a $D=8$ spacetime derivative whereas 
the dilatons also occur without such a derivative.}, $\chi_1,\chi_2$ and 
$\chi_3$. It is useful to define the local $SO(3)$ invariant scalar matrix
\begin{equation}
  \mathcal{M}_{mn} = - L_m{}^i L_n{}^j \eta_{ij} \, ,
\label{Mscalar}
\end{equation} 
where $\eta_{ij} = -\unity_3$ is the internal flat metric. Similarly, the 
two-dimensional $SL(2,\mathbb{R}) / SO(2)$ scalar coset is parameterized by 
the dilaton $\varphi$ and the axion $\ell$ via the local $SO(2)$ invariant scalar 
matrix
\begin{equation}
{\cal W}_{IJ} = e^{\varphi} \left( \begin{array}{cc}
\ell^2 + e^{-2 \varphi} & \ell  \\
\ell    &  1          \\
\end{array}
\right)\, .
\label{Wscalar}
\end{equation}

The only dependence on the internal coordinates $z^m$ comes in
via the $GL(3,\mathbb{R})$ matrices $U^{m}{}_{n}$.  These can be
interpreted as the components of the 3 Maurer-Cartan 1-forms
$\sigma^{m}\equiv U^{m}{}_{n}dz^{n}$ of some 3-dimensional Lie group. 
By definition they satisfy the Maurer-Cartan equations
\begin{equation}
  d\sigma^{m} =-{\textstyle\frac{1}{2}}
  f_{np}{}^{m}\sigma^{n}\wedge \sigma^{p} \,, \qquad
  f_{mn}{}^{p} = -2(U^{-1})^{r}{}_{m} (U^{-1})^{s}{}_{n}\, 
 \partial_{[r} U^{p}{}_{s]}\, ,
\label{MC}
\end{equation}
where the $f_{mn}{}^{p}$ are independent of $z^m$ and form 
the structure constants of the group manifold.  

A subtlety which is not obvious from the analysis by Scherk and Schwarz is 
that only for traceless structure constants ($f_{mn}{}^m = 0 $) one can reduce 
the action \cite{Pons:1998tt}. These cases lead to the class A gauged 
supergravities. For structure constants with non-vanishing trace ($f_{mn}{}^m 
\neq 0 $) one has to resort to a reduction of the field equations. These cases 
lead to the class B gauged supergravities. Note that the embedding of the 
gauge group ${\cal G} \subset GL(3,\mathbb{R})$ is described by
\begin{equation}
g_n{}^m = e^{i\lambda^k f_{kn}{}^m}\, ,
\end{equation}
where $\lambda^k$ are the parameters of the gauge transformations.
Therefore, in the
case of a non-vanishing trace, 
the gauge group ${\cal G}$ is a subgroup of $GL(3,\mathbb{R}) =
SL(3,\mathbb{R}) \otimes SO(1,1)$ and not just $SL(3,\mathbb{R})$.

Using a particular frame in the internal directions,
the explicit coordinate dependence of the 
Maurer-Cartan one-forms 
corresponding to class A is given by 
\begin{align}
  U^m{}_n = \left(
  \begin{array}{ccc}
    1 & 0 & s_{1,3,2} \\
    0 & c_{2,3,1} & - c_{1,3,2} \, s_{2,3,1} \\
    0 & s_{3,2,1} & c_{1,3,2} \, c_{2,3,1}
  \end{array} \right) \,,\hskip 1truecm {\rm det}\ U \ne 1\, ,
\label{explicitU}
\end{align}
where we have used the following abbreviations ($a,b,c=1,2,3$): 
\begin{align}
  c_{a,b,c} \equiv \cos(\sqrt{\tfrac{1}{4} q_a q_b} \, z^c) \,, \qquad
  s_{a,b,c} \equiv \sqrt{q_a/q_b} \sin(\sqrt{\tfrac{1}{4} q_a q_b} \, z^c) \,.
\end{align}
This gives rise to structure constants $f_{mn}{}^p = \epsilon_{mnq}
\mathsf{Q}^{pq}$ with $\mathsf{Q}^{pq} = \tfrac{1}{2}
\text{diag}(q_1,q_2,q_3)$. In Section~\ref{8dsugras} we will explain
that this actually suffices to study all class A gauged supergravities
we obtain.  

Note that the $U$-matrix is independent of $z^3$. It is always possible to 
choose a frame where $z^3$ is a manifest isometry. We distinguish the following three 
different cases:

\begin{description}
\item{(1)}
The matrix $\mathsf{Q}$ is non-singular. In this case $z^3$ is the only manifest
isometry. In the compact case we are dealing with the Salam-Sezgin case 
in which  
the group manifold is equal to $S^3$. The presence of the manifest $z^3$-isometry 
direction is related to the fact that $S^3$ can be viewed as a Hopf fibration 
over $S^2$. One consequence of this fact is that the $D=8$ class A 
supergravities can also be obtained by reduction of the massless IIA theory. 
For instance, the Salam-Sezgin theory can alternatively be obtained by 
reduction of the massless IIA theory over $S^2$. The latter reduction 
naturally occurs in the context of the DW/QFT correspondence 
\cite{Boonstra:1998mp}. In the non-compact case the $SO(3)$ gauging gets 
replaced by an $SO(2,1)$ gauging. This case can be understood as an analytic 
continuation of the Salam-Sezgin theory or as a 
``non-compactification''{}\footnote{As we will see in the next Section in a 
non-compactification we have to discard an infinite factor in front of the 
action. Nevertheless, the procedure leads to a well-defined $D=8$ gauged 
supergravity and, furthermore, can be used as a solution-generating 
transformation of $D=11$ supergravity.} of $D=11$ supergravity.

\item{(2)}
The matrix $\mathsf{Q}$ is singular, e.g. $\mathsf{Q} = \frac{1}{2}{\rm 
diag}(0,q_2,q_3)$. In this case there is an additional isometry in the 
$z^2$-direction:
\begin{align}
  U^m{}_n = \left(
  \begin{array}{ccc}
    1 & 0 & 0 \\
    0 & \cos \alpha & - \sqrt{q_2/q_3}\sin \alpha \\
    0 & \sqrt{q_3/q_2}\sin \alpha & \cos \alpha
  \end{array} \right) \,,\hskip 1truecm {\rm det}\ U = 1\, ,
\label{q2q3}
\end{align}
with $\alpha =\sqrt{\frac{1}{4} q_2 q_3} z^1$. This means that the resulting 
$D=8$ class A gauged supergravities can also be obtained by a reduction of the 
massless 9D theory.

\item{(3)} The matrix $\mathsf{Q}$ is doubly-degenerate, e.g.$\mathsf{Q} = \frac{1}{2}{\rm diag}(0,0,q_3)$. In this case the $U$-matrix is 
given by
\begin{align}
  U^m{}_n = \left(
  \begin{array}{ccc}
    1 & 0 & 0 \\
    0 & 1 & 0 \\
    0 & \frac{1}{2}q_3 z^1 & 1
  \end{array} \right) \,,\hskip 1truecm {\rm det}\ U = 1\, ,
\label{q3}
\end{align}
and again the resulting $D=8$ class A gauged supergravity has its origin in 
the massless 9D theory.

\end{description}

The g.c.t.'s in the internal space generate symmetries in $D=8$
dimensions.  These g.c.t.'s are generated by the Killing vector
\begin{equation}
\hat{K}^{m} (\hat{x}) = -(U^{-1})^{m}{}_{n}(z) \lambda^{n}(x) +R^{m}{}_{n}z^{n} \,. 
\end{equation}
Upon reduction these correspond to 
\begin{itemize}
\item $GL(3,\mathbb{R})$ transformations with parameters
  $R^{m}{}_{n}$. These can be decomposed into $SL(3,\mathbb{R})$
  rotations, that act in the obvious way on all the fields that have
  $m,n,p$ indices, and $SO(1,1)$ rescalings.
\item Gauge transformations with parameters $\lambda^{m}$ or
$\lambda_{m}{}^{n}\equiv -f_{mp}{}^{n}\lambda^{p}$
that act on all the fields that have $m,n,p$ indices e.g.~
\begin{equation}
\delta_{\lambda} L_{m}{}^{i} 
= -L_{n}{}^{i} \lambda_{m}{}^{n}=
f_{mp}{}^{n} \lambda^{p} L_{n}{}^{i}\, , 
\label{8dgaugetransf}
\end{equation}
except for the Kaluza-Klein vectors $A^{m}{}_{\mu}$ that transform as gauge 
vectors,
\begin{equation}
\delta_{\lambda} A^{m}{}_{\mu} = {\cal D}_{\mu} \lambda^{m}
                                 \equiv\partial_{\mu} \lambda^{m}
                              -f_{np}{}^{m} A^{n}{}_{\mu} \lambda^{p} \, ,
\end{equation}
of the gauge group $\cal G$.
\end{itemize}

Performing the reduction of the 11-dimensional bosonic action
\eqref{eq:11dtruncaction} and supersymmetry variation of the gravitino
\eqref{eq:11Dsusy} with the above reduction Ansatz, restricted to traceless
structure constants, we find the class A $D=8$ gauged supergravities described 
in the next Section. Although in principle straightforward, we will not 
perform the explicit reduction for the class B theories in this paper.

\section{Gauged Supergravities in $D=8$} \label{8dsugras}

As discussed in the previous Section, the different gauged supergravities can 
be obtained by a reduction of the 11-dimensional supergravity over different 
group manifolds. We restrict ourselves to gauge groups with traceless 
structure constants: $f_{mn}{}^n = 0$.
For simplicity, we will only reduce the bosonic part of the action and consider the supersymmetry 
rules up to bilinears in the fermions. We give the full bosonic 8D action in Appendix 
\ref{full8daction}. Here we consider the truncation that the $D=11$ three-form 
potential is equal to zero. In this truncation the reduction of the 11D 
Einstein-Hilbert term gives rise to
\begin{equation}
\label{eq:8dtruncaction}
S  =  
{\textstyle\frac{1}{16\pi G_{N}^{(11)}}}\, C_U
{\displaystyle\int} d^{8}x \sqrt{|g|}\,
 \big[ R
+{\textstyle\frac{1}{4}}{\rm Tr}\left({\cal D} {\cal M}{\cal M}^{-1}\right)^{2}
+{\textstyle\frac{1}{2}}(\partial \varphi)^2 
- {\textstyle\frac{1}{4}}e^{-\varphi}F^{m}{\cal M}_{mn} F^{n}
- {\cal V} \big] \, ,
\end{equation}
where the $SL(3,\mathbb{R}) / SO(3)$ scalar matrix $\cal M$ is defined in
\eqref{Mscalar}, the potential ${\cal V}$ is given by
\begin{equation}
 {\cal V} ={\textstyle \frac{1}{4}}e^{-\varphi}\,
             \left[ 2{\cal M}^{nq}f_{mn}{}^{p} f_{pq}{}^{m}
                  + {\cal M}^{mq}{\cal M}^{nr}{\cal M}_{ps}
                  f_{mn}{}^{p}f_{qr}{}^{s}
             \right]\, 
\end{equation}
and $C_U$ is defined by
\begin{align}
  C_U = \int dz^m {\rm det}\, (U^{m}{}_{n}) \,.
\end{align}
The integral that defines the factor $C_{U}$ generally converges. It only 
diverges in the non-compact version of the Salam-Sezgin theory. The resulting 
$D=8$ $SO(2,1)$-gauged supergravity action is a ``non-compactification'' of 
$D=11$ supergravity, see also the discussion in the previous Section.

The covariant derivative $\cal D$ is always with respect to the gauge group $\cal G$ 
defined by the structure constants $f_{mn}{}^p$ and the gauge vectors $A^{m}$. Thus
the gauge vector field strengths and the covariant derivative of the scalar coset read
\begin{equation}
  F^{m} = 2\partial A^{m} -f_{np}{}^{m}A^{n}A^{p} \,, \qquad
  {\cal D} {\cal M}_{mn} = \partial {\cal M}_{mn} + 2 f_{q(m}{}^{p} A^{q} {\cal M}_{n)p} \,.
\end{equation}

The supersymmetry variations of the fermions read (in the truncation
we are considering here)
\begin{align}
\delta \psi_\mu & = 2 \partial_\mu \epsilon - \tfrac{1}{2} \slashed{\omega}_\mu \epsilon  
   +\frac{1}{2} \slashed{Q}_\mu \epsilon 
     -\tfrac{1}{2} e^{\varphi/2} \Gamma^m
  ( \tfrac{1}{12}\Gamma_{\mu}{\slashed F}_m + 
  F_{m \mu \nu} \Gamma^\nu )  \epsilon 
 + \tfrac{1}{24} e^{- \varphi /2} f_{ijk} \Gamma^{ijk} 
\Gamma_\mu \epsilon \,, \notag \\
\delta \lambda_i & =  
 - \slashed P_{ij}\Gamma^{j}\epsilon 
  - \tfrac{1}{3} \slashed{\partial} \varphi \Gamma_i \epsilon 
  + \tfrac{1}{4}e^{\varphi/2}{\cal M}_{in}{\slashed F}^n \epsilon 
  - \tfrac{1}{4} e^{-\varphi/2} (2 f_{ijk} - f_{jki}) \Gamma^{jk} 
\epsilon  \,,
\end{align}
where we have used the abbreviations $f_{ijk} \equiv L_i{}^m L_j{}^n L_{pk} f_{mn}{}^p$ and
\begin{align}
  P_\mu{}_{ij} + Q_\mu{}_{ij} & \equiv L_i{}^m {\cal D}_\mu L_{mj} \,, \qquad
  \slashed P_{ij} \equiv P_\mu{}_{ij} \Gamma^\mu \,, \qquad
  \slashed{Q}_\mu \equiv Q_\mu{}_{ij} \Gamma^{ij} \,,
\end{align}
where $P$ is symmetric and traceless and $Q$ is antisymmetric.

We observe that the massive deformations $f_{mn}{}^p$ come from the
reduction over the group manifold.  The choice $f_{mn}{}^p=0$ is
the ungauged case and corresponds to reduction over $T^3$ leading to
the trivial gauge group $U(1)^3$.  The full
supergravity theory has a global $SL(3,\mathbb{R}) \times
SL(2,\mathbb{R})$ symmetry group in the massless case.  The
$SL(3,\mathbb{R})$ symmetry acts in the obvious way on the indices
$m,n$, while the $SL(2,\mathbb{R})$ symmetry is not a manifest
symmetry of the action.  Choosing non-vanishing structure constants
modifies this symmetry group in the following way.  The
$SL(2,\mathbb{R})$ symmetry is fully broken, while the
$SL(3,\mathbb{R})$ symmetry generically is broken due to the structure
constants.  Performing an $SL(3,\mathbb{R})$ transformation has the
effect of changing the structure constants via
\begin{align}
  f_{mn}{}^p \rightarrow f'_{mn}{}^p = R_m{}^q R_n{}^r 
(R^{-1})_s{}^p f_{qr}{}^s \,.
\label{structtransf}
\end{align}
Only transformations that leave the structure constants invariant
($f_{mn}{}^p = f'_{mn}{}^p$) are unbroken by the massive
deformations.  This includes the infinitesimal gauge transformations
\eqref{8dgaugetransf}.

The structure constants of all 3-dimensional Lie algebras can be parameterized 
by a symmetric matrix that we denote by $\mathsf{Q}^{mn}$ and which will play 
the role of mass matrix, and by a vector $a_{m}$ satisfying 
$\mathsf{Q}^{mn}a_{n}=0$ (see, e.g., \cite{Wald:1984rg}):
\begin{equation}
 f_{mn}{}^{p}= \epsilon_{mnq} \mathsf{Q}^{qp} +2\delta_{[m}{}^{p}a_{n]}\, . 
\end{equation}
The trace of the structure constants vanishes if and only if the
vector vanishes, i.e.~$a_{m}=0$. Restricting to the class A gauged
supergravities we can take $f_{mn}{}^{p}= \epsilon_{mnq} \mathsf{Q}^{qp}$
and all the different cases that we are going to consider will be
characterized by a choice of mass matrix $\mathsf{Q}$.

In terms of the mass matrix the potential reads
\begin{align}
 {\cal V} = - \tfrac{1}{2} \, e^{-\varphi}\, \{ \left[ {\rm Tr}({\cal M}{\mathsf Q}) \right]^2
   - 2 {\rm Tr}({\cal M}{\mathsf Q}{\cal M}{\mathsf Q})\}\, .
\label{potentialQ}
\end{align}
The symmetric mass matrix $\mathsf{Q}$  has six different mass parameters.
However, by applying symmetries of the massless 8D theory one can
relate different choices of $\mathsf{Q}^{mn}$ by field redefinitions,
via transformations as \eqref{structtransf}.  We would like to use the
$GL(3,\mathbb{R})$ symmetry of the massless 8D theory.

Employing these symmetries we can transform $\mathsf{Q}^{mn}
\rightarrow \pm (R^T \mathsf{Q} R)^{mn}$ with $R \in
GL(3,\mathbb{R})$.  Now consider an arbitrary symmetric matrix
$\mathsf{Q}^{mn}$ with eigenvalues $\lambda_m$ and orthogonal
eigenvectors $\vec{u}_m$.  Taking $R = (c_1 \vec{u}_1, c_2
\vec{u}_2, c_3 \vec{u}_3) \in GL(3,\mathbb{R})$ with $c_i \neq 0$ we
find that
\begin{align}
  \mathsf{Q}^{mn} \rightarrow \pm (R^T \mathsf{Q} R)^{mn} = 
\pm \rm{diag}(c_1{}^2 \lambda_1, c_2{}^2 \lambda_2, c_3{}^2 \lambda_3)\, ,
\end{align}
which is a minor extension of the Principal Axes theorem.  Thus
all cases with the same signature are related by field redefinitions.
Without loss of generality we will use the freedom of field redefinitions 
to take
\begin{align}
  \mathsf{Q}^{mn} = \tfrac{1}{2} {\rm diag}(q_1,q_2,q_3) \, .
\end{align}
The different 8D massive supergravities will arise from choosing all possible
ranks and signatures for the mass matrix $\mathsf{Q}^{mn}$.

\begin{table}[ht]
\begin{center}
\begin{tabular}{||c|c|c||}
\hline \rule[-3mm]{0mm}{8mm}
  Bianchi & $\mathsf{Q}=\tfrac{1}{2} \text{diag}$ & Group \\
\hline \hline \rule[-3mm]{0mm}{8mm}
  II & $(0,0,q)$ & Heisenberg \\
\hline \rule[-3mm]{0mm}{8mm} 
  VI$_0$ & $(0,-q,q)$ & $ISO(1,1)$ \\
\hline \rule[-3mm]{0mm}{8mm} 
  VII$_0$ & $(0,q,q)$ & $ISO(2)$ \\
\hline \rule[-3mm]{0mm}{8mm} 
  VIII & $(q,-q,q)$ & $SO(2,1)$ \\
\hline \rule[-3mm]{0mm}{8mm} 
  IX & $(q,q,q)$ & $SO(3)$ \\
\hline
\end{tabular}
\caption{\it The different mass matrices and corresponding Bianchi classifications and gauge groups. 
The $SO(3)$ result was previously obtained in \cite{Salam:1985ft}. 
\label{gaugegroups}}
\end{center}
\end{table}

Actually, this diagonalization plus the choice $a_{m}=(a,0,0)$ is
the basis of the Bianchi classification of all real 3-dimensional Lie
algebras from Bianchi type~I to Bianchi type~IX. Thus, each choice of
mass matrix corresponds to a choice of Lie algebra and therefore of
gauge group. Restricting to the class A theories we only consider the
algebras with $a=0${}\footnote{The sub-index $0$ in Bianchi type~${\rm
    VI_0}$ and Bianchi type~${\rm VII_0}$ indicate that these class A
  Lie algebras can be obtained as the limit $a \rightarrow 0$ of the
  class B Bianchi type~${\rm VI}_a$ and Bianchi type~${\rm VII}_{a}$
  Lie algebras (see appendix D).}:
\begin{equation}
  \text{Bianchi types I, II, VI$_0$, VII$_0$, VIII, IX}\, .
\end{equation}
All algebras with $a=0$ are subalgebras of the
Lie algebra of $SL(3,\mathbb{R})$. For useful details about the
Bianchi classification, see appendix D.
The five nontrivial cases with $a=0$ are given in Table \ref{gaugegroups}
while Bianchi type~I corresponds to the massless case $\mathsf{Q}=0$ and thus
is an ungauged supergravity. This case corresponds to the Abelian 
Lie algebra $U(1)^3$.

\section{The Domain Wall Solutions} \label{DWsolutions}

Having obtained the bosonic action and supersymmetry transformations of the 
$D=8$ gauged maximal supergravities with gauge groups of class A, we now look 
for domain wall solutions that preserve half of the supersymmetry. For an 
earlier discussion of such solutions, see 
\cite{Cowdall:1997tw,Hernandez:2002fb}. We consider the following Ansatz:
\begin{align}
  ds^2 & = g(y)^2 dx_7{}^2 - f(y)^2 dy^2 \,, \qquad
{\cal M} = {\cal M}(y)\, ,\hskip .6truecm \varphi = \varphi (y)\,, \qquad
  \epsilon = \epsilon(y) \,.
\label{Ansatz}
\end{align}
Our Ansatz only includes the metric and the scalars. 
All other fields are vanishing except the $SL(2,\mathbb{R}) / SO(2)$ scalar $\ell$
which we have set constant. It turns out that there 
are no half-supersymmetric domain walls for non--constant $\ell$.

We need to satisfy the Killing spinor equations
\begin{align}
\delta \psi_\mu & = 2 \partial_\mu \epsilon 
-\tfrac{1}{2} \slashed{\omega}_\mu \epsilon  
   +\tfrac{1}{2} \slashed{Q}_\mu \epsilon 
   + \tfrac{1}{24} e^{- \varphi /2} f_{ijk} \Gamma^{ijk} 
\Gamma_\mu \epsilon = 0 \, , \notag \\
\delta \lambda_i & = - \slashed P_{ij}\Gamma^{j}\epsilon 
  - \tfrac{1}{3} \slashed{\partial} \varphi \Gamma_i \epsilon 
  - \tfrac{1}{4} e^{-\varphi/2} (2 f_{ijk} 
-f_{jki}) \Gamma^{jk} \epsilon = 0 \, , \notag 
\end{align}
where the Killing spinor satisfies the condition
\begin{align}
  (1+ \Gamma_{y123}) \epsilon = 0 \, .
\end{align}
The indices $1,2,3$ refer to the internal group manifold
directions. 

The domain wall solutions we present below are valid both for a
non-singular and a singular mass matrix $\mathsf{Q}$. We
find the following most general class A solution:
\begin{align}\label{triple}
  ds^2 & = H^{\frac{1}{12}}dx_7^2-H^{-\frac{5}{12}}dy^2\,, \nonumber \\
  e^\varphi & = H^{\frac{1}{4}}, \hspace{0.5cm} 
  e^\sigma = H^{-\frac{1}{2\sqrt{3}}}h_1^{\frac{\sqrt{3}}{2}}, \hspace{0.5cm}
  e^\phi= H^{-\frac{1}{2}}h_1^{-\frac{1}{2}}(h_1h_2-C_1^2)\,, \notag \\
  \chi_1 & = C_1h_1^{-1}, \hspace{0.5cm} 
  \chi_2=\chi_1\chi_3+C_2h_1^{-1},\hspace{0.5cm}
  \chi_3=(C_1C_2+C_3 h_1)\left(h_1h_2-C_1^2 \right )^{-1},
\end{align}
where the dependence on the transverse coordinate $y$ is governed by
\begin{align}
& H(y) = h_1 h_2 h_3 - C_3^2 h_1-C_2^2 h_2-C_1^2 h_3-2C_1 C_2 C_3 \,, \notag \\
& h_1 \equiv q_1y+C_4, \hspace{0.5cm} h_2\equiv q_2y+C_5, \hspace{0.5cm} 
h_3\equiv q_3y+C_6\,.
\label{harmonics}
\end{align}
The corresponding Killing spinor is quite intricate so we will not give it here.
Note that the solution is given by three harmonic function $h_1$, $h_2$ and 
$h_3$. For this reason we call the general solution a triple domain wall.

The general solution has six integration constants $C_1, \ldots , C_6$. Note 
that the constants $q_1, q_2$ and $q_3$ should not be considered as parameters 
of the solution. They only serve to specify which of the five supergravities 
we are dealing with. The constants $C_4$, $C_5$ and $C_6$ are related to the 
positioning of the domain walls in the transverse space and $q_1$, $q_2$ and 
$q_3$ are their respective charges. The domain walls form a threshold bound 
state of $n$ parallel domain walls, where $n$ equals the rank of the mass 
matrix. It turns out that, provided that one of the charges $q_1, q_2$ or 
$q_3$ is non-zero, one can eliminate one of the constants $C_4, C_5$ or $C_6$ 
by a redefinition of the variable $y$. Therefore we effectively always end up 
with at most two constants.

The first three constants $C_1$, $C_2$ and $C_3$ can be understood to come 
from the following symmetry. The mass deformations do not break the full 
global $SL(3,\mathbb{R})$; indeed, they gauge the 3-dimensional subgroup of 
$SL(3,\mathbb{R})$ that leave the mass matrix $\mathsf{Q}$ invariant. Thus one 
can use the unbroken global subgroup to transform any solution\footnote{
  Note that one can not use the unbroken local subgroup of $SL(3,\mathbb{R})$ 
  (the gauge transformations) since this would induce non-vanishing
  gauge vectors and thus would be inconsistent with our Ansatz \eqref{Ansatz}.}
, introducing 
three constants. In our solution these correspond to $C_1$, $C_2$ and $C_3$ 
and thus these can be set to zero by fixing the $SL(3,\mathbb{R})$ frame. From 
now on we will always assume the frame choice $C_1 = C_2 = C_3 = 0$ unless 
explicitly stated otherwise. This results in
\begin{align}\label{basis}
  \chi_1 = \chi_2 = \chi_3 = 0 \,, \qquad 
  \mathcal{M} = H^{-2/3} \text{diag}( h_2 h_3, h_1 h_3, h_1 h_2 ) \,, \qquad 
  H = h_1 h_2 h_3 \,.
\end{align}
This structure is very similar to that found in \cite{Bakas:1999fa}, upon 
which we will comment in the Conclusions. In this $SL(3,\mathbb{R})$ frame the
expression for the Killing spinor simplifies considerably and reads
$\epsilon = H^{1/48} \epsilon_0$.

The triple domain wall can be truncated to double or single domain walls when 
restricting the constants $C_4, C_5$ and $C_6$. The single domain walls 
correspond to the situation where the positions of the parallel domain walls 
coincide. In Table \ref{singleDW} we give the three possible truncations 
leading to single domain walls and the corresponding value of $\Delta$ as 
defined in \cite{Lu:1995cs}. The Bianchi II case was given in 
\cite{Cowdall:1997tw} and the Bianchi IX case in \cite{Boonstra:1998mp} (up to 
coordinate transformations). Note that the Bianchi VII$_0$ case can not be 
assigned a $\Delta$-value since it has vanishing potential. The domain wall is 
carried by the non-vanishing massive contributions to the BPS equations. The 
same mechanism occurs in $SO(2)$ gauged $D=9$ supergravity 
\cite{Bergshoeff:2002mb}.

\begin{table}[ht]
\begin{center}
\begin{tabular}{||c||c|c|c|c||c||c||}
\hline \rule[-3mm]{0mm}{8mm}
  Bianchi & $\mathsf{Q}=\tfrac{1}{2} \text{diag}$ & $h_1$ & $h_2$ & $h_3$ & $\Delta$ & Uplift \\
\hline \hline \rule[-3mm]{0mm}{8mm}
  II & $(0,0,q)$ & $C_4$ & $C_5$ & $C + q y$ & $4$ & \eqref{4EH-II-Heis} \\
\hline \rule[-3mm]{0mm}{8mm} 
  VII$_0$ & $(0,q,q)$ & $C_4$ & $C + q y$ & $C +q y$ & $\times$ & \eqref{4EH-I-ISO}  \\
\hline \rule[-3mm]{0mm}{8mm} 
  IX & $(q,q,q)$ & $C+qy$ & $C + q y$ & $C + q y$ & $-\tfrac{4}{3}$ & \eqref {4flat} \\
\hline
\end{tabular}
\caption{\it The single domain walls as truncations of the triple domain wall 
solution. We give the three possible truncations and the corresponding value 
of $\Delta$ \cite{Lu:1995cs}. Note that there exists no $\Delta$-value in the 
Bianchi VII$_0$ case due to the vanishing of the potential. In the last column 
we indicate the equation where the uplifted solution to 11D is given. \label{singleDW}}
\end{center}
\end{table}

The triple domain wall solution we found in this Section can be interpreted as 
follows. One can view the $(0,0,q)$ solution, having one harmonic function, as 
the basic solution. The other solutions can then be obtained as threshold 
bound states of this solution with the $SL(3,\mathbb{R})$-rotated solutions 
$(0,q,0)$ and $(q,0,0)$. This is clear at the level of the charges. We now see 
how, similarly, a composition rule at the level of the solutions can be 
established. One can thus view the solutions with a rank-1 mass matrix as 
building blocks for the general solution.

\section{Uplifting to 11 Dimensions}

In this Section we consider the uplifting of the triple domain wall solutions 
(\ref{triple}) to eleven dimensions. We find that upon uplifting, using the 
frame of (\ref{basis}), the triple domain wall solutions becomes a purely 
gravitational solutions with a metric of the form $\hat{ds}{}^2 
=dx_7{}^2-ds_4{}^2$, where
\begin{align}
  ds_4{}^2 = H^{-\frac{1}{2}}dy^2+H^{\frac{1}{2}}
\left ( \frac{\sigma_1^2}{h_1}+\frac{\sigma_2^2}{h_2}+\frac{\sigma_3^2}{h_3} 
\right) \,. \label{4dmetric}
\end{align}
Here $\sigma_1$, $\sigma_2$ and $\sigma_3$ are the Maurer-Cartan 1-forms 
defined in \eqref{MC}, $H=h_1h_2h_3$ and the three harmonics $h_1,h_2$ and 
$h_3$ are defined in \eqref{harmonics}. The uplifted solutions are all 1/2 BPS.

The solutions (\ref{4dmetric}) are cohomogeneity one solutions of different 
Bianchi types. The $SO(3)$ expression of this 4-dimensional metric was 
obtained previously in the context of gravitational instanton solutions as 
self-dual metrics of Bianchi IX type with all directions unequal 
\cite{Belinskii:1978}. We find that the metric of \cite{Belinskii:1978} is 
related to a triple domain wall solution of 8-dimensional $SO(3)$-gauged 
supergravity. More recently, the Heisenberg, $ISO(1,1)$ and $ISO(2)$ cases and 
their relations to domain wall solutions were considered in 
\cite{Gibbons:1998ie,Lavrinenko:1998qa}, whose results are related to ours via
coordinate transformations.

It is remarkable that in all cases the uplifted solutions have metrics that 
contain a 7D Minkowski metric as a factor \cite{Hernandez:2002fb}. This does 
not happen for the uplift of domain walls in 4D and 7D gauged maximal 
supergravities \cite{Bakas:1999fa}. In the following discussion we will focus 
on the 4-dimensional part of the eleven-dimensional metric since it 
characterizes the uplifted domain walls.

In Section \ref{DWsolutions} we argued that for $q_1$, $q_2$ or $q_3$ non-zero 
one of the three constants $C_4$, $C_5$ or $C_6$ in the harmonic functions can be 
eliminated by a redefinition of the variable $y$. Without loss of generality 
we can take $q_3>0$. In that case the constant $C_6$ can be eliminated by the 
change of variables $y = \tfrac{1}{4}r^4- C_6/q_3$ in the metric 
(\ref{4dmetric}). If we also rescale the three charges by $q_m = 4 \tilde q_m$ 
we obtain
\begin{align}\label{2pardmetric}
  ds_4{}^2 & = (k_1 k_2k_3)^{-1/2} \left[ dr^2+ 
  r^2  ( k_2 k_3 \sigma_1^2 + k_1k_3 \sigma_2^2 + k_1 k_2 \sigma_3^2 )
  \right] \, ,
\end{align}
where $k_i = \tilde q_i + s_i r^{-4}$ for $i=1,2$, $s_i=C_{i+3}-q_iC_6/q_3$ 
and $k_3= \tilde q_3$. As anticipated, the metric 
(\ref{2pardmetric}) depends only on two constant parameters $s_1$ and $s_2$, 
which are restricted by the gauge group dependent condition $-s_i< \tilde q_i 
r^4$, with $i=1,2$, in order to satisfy the requirement $k_i(r)> 0$. 

In general the metrics (\ref{2pardmetric}) have 
curvatures that both go to zero as $r^{-6}$ for large $r$ and diverge at 
$r=0$, $r= (-s_1/\tilde q_1)^{1/4}$ and $r= (-s_2/\tilde q_2)^{1/4}$, producing incomplete metrics 
\cite{Eguchi:1978,Belinskii:1978}. There are two exceptions to this behavior. 
The first one corresponds to the case in which $s_1=s_2=0$. The constants 
can take these values because $\tilde q_1$ and $\tilde q_2$ are non-zero and 
therefore this solution can be reached only for the non-degenerate cases. It 
is easy to see that for the $SO(3)$ gauging ($\tilde q_1=\tilde q_2 =\tilde 
q_3=1$) the metric is locally flat space-time
\begin{align}\label{4flat}
  ds_4{}^2 & = dr^2+ r^2 ((\sigma^1)^2+(\sigma^2)^2+(\sigma^3)^2) \, ,
\end{align}
where $r$ is the radius of the 3-dimensional spheres. Notice that this is 
precisely the uplifting of the Bianchi IX single domain wall. This should 
correspond to the orbifold limit of the K3 manifold and therefore it is still 
only 1/2 BPS upon uplifting.

The second exception corresponds to the $SO(3)$ gauging with $s_1=s_2 \equiv s 
<0$, and is known as the Eguchi-Hanson (EH), or Eguchi-Hanson II, metric 
\cite{Eguchi:1978} ($\tilde q_1=\tilde q_2 =\tilde 
q_3=1$),
\begin{align}\label{4EH-II}
  ds_4{}^2 & = \left( 1+\frac{s}{r^4} \right)^{-1}dr^2
+ r^2 (\sigma_1^2+\sigma_2^2) +\left( 1 +\frac{s}{r^4} \right)\sigma_3^2 \, .
\end{align}
In fact, the EH metric is the only complete and non-singular hyper-K\"ahler 
4-metric admitting a tri-holomorphic $SO(3)$ action. Its generic orbits are 
$RP^3$ \cite{Belinskii:1978,Gibbons:1979,Gibbons:2002}.

Another case that we want to emphasize, although it is singular, is obtained 
in the $SO(3)$ gauging by choosing $s_1\equiv s \neq 0$ and $s_2 = 0$. This 
metric is called the Eguchi-Hanson I (EH-I) metric \cite{Eguchi:1978}
($\tilde q_1=\tilde q_2 =\tilde 
q_3=1$)
\begin{align}\label{4EH-I}
  ds_4{}^2 & = \left( 1+\frac{s}{r^4} \right)^{-1}(dr^2
+ r^2 \sigma_1^2) +\left( 1 +\frac{s}{r^4} \right)(\sigma_2^2+\sigma_3^2) \, .
\end{align}
It is possible to give similar expressions for the $SO(2,1)$ gauging.

The uplifted metrics for the singular mass matrices can be
 obtained directly from 
(\ref{2pardmetric}).
As an example of a contraction we take $\tilde q_1=0$ in 
(\ref{2pardmetric}) and consider the special cases of the
EH metrics 
(\ref{4EH-II}) and (\ref{4EH-I}), i.e.~we take 
$s_1 = s_2 \equiv s <0$ (EH-II) or $s_1 \equiv s \ne 0, s_2 = 0$ (EH-I).
We thus obtain the contracted EH metrics with 
$ISO(2)$ isometry in which the $SO(3)$ orbits are flattened to $ISO(2)$ 
orbits. We find that the expression for the contracted EH-I metric is
given by ($\tilde q_2 =\tilde q_3=1$)
\begin{align}\label{4EH-I-ISO}
  ds_4{}^2 & = \left(\frac{s}{r^4} \right)^{-1/2}(dr^2
+ r^2 \sigma_1^2) +\left(\frac{s}{r^4} \right)^{1/2}(\sigma_2^2+\sigma_3^2) \,,
\end{align}
while the expression for the contracted EH-II metric reads
($\tilde q_2 =\tilde q_3=1$)
\begin{align}\label{4EH-II-ISO}
  ds_4{}^2 = & \left(\frac{s}{r^4}(1+\frac{s}{r^4}) \right)^{-1/2}dr^2
+ \left( \frac{s}{r^4}(1 +\frac{s}{r^4}) \right)^{1/2}\sigma_3^2\ +\nonumber \\ 
& + r^2 \Big( ( 1+\frac{r^4}{s} )^{1/2}\sigma_1^2+ ( 
1+\frac{r^4}{s} )^{-1/2}\sigma_2^2 \Big) \, .
\end{align}
Notice that the 
contracted EH-I metric with $ISO(2)$ isometry is precisely the 4-dimensional 
part of the uplifted metric for the Bianchi VII$_0$ single domain wall.

The metrics with Heisenberg isometry are obtained by a further contraction 
$\tilde q_1 = \tilde q_2=0$ in the metric (\ref{2pardmetric}). Again, among these metrics 
there is one special case that can also be obtained by a contraction of the 
contracted EH metric with isometry $ISO(2)$. Notice that it is not possible to 
have a contracted EH-I metric with Heisenberg isometry since 
we must satisfy the condition $-s_i < \tilde q_i r^4$.
The expression for the contracted EH metric with Heisenberg isometry is
($\tilde q_3=1$)
\begin{align}\label{4EH-II-Heis}
  ds_4{}^2 & = \left(\frac{s}{r^4} \right)^{-1}dr^2
+ r^2 (\sigma_1^2+\sigma_2^2) +\left(\frac{s}{r^4} \right)\sigma_3^2 \,, \notag \\
  & = \left(\frac{s}{r^4} \right)^{-1}dr^2
+ r^2 (dz_1{}^2+dz_2{}^2) +\left(\frac{s}{r^4} \right)(dz_3+ 2 z^1 d z_2)^2 \,,
\end{align}
where $s_2 \equiv s$. This is the 4-dimensional part of the 
uplifted metric for the Bianchi II single domain wall 
This contraction was considered in \cite{Gibbons:2002}.

\section{Conclusions}

In this paper we have considered two classes, A and B, of $D=8$ gauged maximal 
supergravity theories. Class A contains supergravities with an action while 
the supergravities of class B only have equations of motion. Class A contains 
5 supergravities corresponding to the following five different subgroups of 
$SL(3,\mathbb{R})$: $SO(3), SO(2,1), ISO(2), ISO(1,1)$ and the Heisenberg 
subgroup. We have constructed a general half-supersymmetric triple domain-wall 
solution to these theories. It can be viewed as a threshold bound state of 
three parallel single domain walls. The uplifting of this solution to $D=11$ 
dimensions leads to a purely gravitational solution whose metric is the direct 
product of a 7-dimensional Minkowski metric and a non-trivial 4-dimensional 
Euclidean Ricci-flat metric \cite{Hernandez:2002fb}. The 4-metrics are 
solutions of 4-dimensional Euclidean gravity. Among them we find 
generalizations of the Eguchi-Hanson solution to different (class A) Bianchi 
types.

The results of this paper are similar to the $D=9$ case 
\cite{Bergshoeff:2002mb}: in both cases a $GL(11-D,\mathbb{R})$ group ($D=8, 
9$) and its subgroups are the main characters. The group $GL(11-D,\mathbb{R})$ 
appears naturally in (ungauged) maximal supergravities in $D$ dimensions as 
part of its duality group since they can be obtained by toroidal 
compactification of 11-dimensional supergravity. It is natural to expect the 
existence of gauged supergravities associated to the subgroups of 
$GL(11-D,\mathbb{R})$. Some cases are already well known, for instance the 
$D=5$ maximal supergravities with gauge groups $SO(6-l,l)$ (all of them 
subgroups of $SL(6,\mathbb{R})$) constructed in 
\cite{Gunaydin:1984qu,Pernici:ju}{}\footnote{Other theories in different 
dimensions can also be constructed with an explicit symmetric mass matrix 
$\mathsf{Q}$ present \cite{Alonso-Alberca:2002tb}.}. Gauged maximal
supergravities in diverse dimensions have recently been investigated 
from a somewhat different point of view in \cite{deWit:2002vt}.

An interesting outcome of our analysis is the existence,
in $D=8$ dimensions, of the generic triple 
domain wall solution \eqref{triple}. It can be interpreted as $m$ parallel 
single domain walls where $m$ is the rank of the mass matrix. For the gauging 
of $SO(3)$, this result is similar to that of \cite{Bakas:1999fa}. There the 
scalar content was the coset $SL(n,\mathbb{R}) / SO(n)$ while we have the 
product of $n=3$ with an additional $n=2$ coset. Note that, in the gauged cases, 
our coset can not be reduced to the $SL(3,\mathbb{R}) / SO(3)$ by truncation of the 
$SL(2,\mathbb{R}) / SO(2)$ scalars. It is interesting that the 
structure of \cite{Bakas:1999fa} extends to more general scalar contents and 
to the other Bianchi classes. It leads one to expect a similar $n$-tuple 
domain wall result in other dimensions. In fact, we verified that in $D=9$ 
with scalar content $SO(1,1) \times SL(2,\mathbb{R}) / SO(2)$ the earlier 
results on domain wall solutions in gauged $D=9$ supergravity 
\cite{Bergshoeff:2002mb} can be written as a generic double domain wall 
solution via coordinate transformations.

The relation between $D=8$ domain wall solutions and gauged supergravities 
that we have discussed fits naturally in the domain wall/QFT correspondence 
scheme \cite{Itzhaki:1998dd,Boonstra:1998mp}. As discussed in 
\cite{Boonstra:1998mp}, taking the near-horizon limit of the D6-brane leads to 
the $D=8$ $SO(3)$ gauged supergravity. Taking the near-horizon limit of the 
direct reduction of the D6-brane to $D=9$ dimensions leads to the $D=8$ 
$ISO(2)$-gauged supergravity. A further direct reduction to a 6-brane in $D=8$ 
dimensions leads to the $D=8$ Heisenberg gauged supergravity.

In this paper we mainly concentrated on the construction of the $D=8$ class A 
gauged supergravities. We plan to investigate the class B theories as well. 
One difference with the class A theories is that the Maurer-Cartan 1-forms for
traceful structure constants probably have no additional isometry. Therefore, 
in contrast to the class A case, these reductions cannot be reproduced by any 
known reduction of the massless IIA theory. Cohomogeneity one solutions of 
class B Bianchi type have been considered in the literature 
\cite{Gibbons:1998ie}. It would be interesting to see whether these solutions 
can be reduced to 1/2 BPS domain wall solutions of the corresponding class B 
$D=8$ gauged supergravity.

It is interesting to note that the uplifting of the triple domain wall
solution (\ref{triple}) does not lead to the most general 4-metric with
$SO(3)$ isometry. The complete non-singular $SO(3)$-invariant hyper-K\"ahler 
metrics in four dimensions are the Eguchi-Hanson, Taub-NUT and Atiyah-Hitchin 
metrics (for a useful discussion of these metrics see \cite{Bakas:1997gf}). 
The absence of the Taub-NUT and Atiyah-Hitchin metrics in our analysis is 
related to the fact that only the (generalized) Eguchi-Hanson metric allows a 
covariantly constant spinor that is independent of the $SO(3)$ isometry 
directions \cite{Gauntlett:1997pk}. In performing the SS2 reduction we have 
assumed that our spinors are independent of the group manifold coordinates and 
this assumption is thus not compatible with the Taub-NUT and the 
Atiyah-Hitchin metrics. It would be interesting to see whether we can relax 
the SS2 procedure such that the Taub-NUT and Atiyah-Hitchin metrics also 
obtain a half-supersymmetric domain wall interpretation in $D=8$ dimensions or 
whether we should view them as $D=8$ domain walls with fully broken 
supersymmetry.

In the same spirit one can hope to extend the SS1 reduction, for example as 
applied in \cite{Bergshoeff:2002mb}. In that paper the spinors generally were 
transforming under the $SL(2,\mathbb{R})$ duality symmetry and, consequently, 
the spinors were given dependence on the internal direction. However, for 
contracted group manifolds, our Ans\"atze with dependence on $z_1$ only, 
see \eqref{q2q3},
can be interpreted as a reduction from the massless 9D theory. In 
this case we have taken the spinors to be independent of the internal 
direction. We therefore have two reduction Ans\"atze that only differ in the fermionic sector.
Therefore, both the SS1 and SS2 reduction procedure might be amenable to extension 
and it would be 
desirable to understand the differences between the resulting 
gauged supergravities.

\medskip
\section*{Acknowledgments}

\noindent 
T.O.~would like to thank the CERN Theory Division for its hospitality
and financial support and M.M.~Fern\'andez for her continuous support.
This work is supported in part by the European Community's Human
Potential Programme under contract HPRN-CT-2000-00131 Quantum
Spacetime, in which the University of Groningen is associated with the
University of Utrecht.  The work of U.G. is part of the
research program of the ``Stichting voor Fundamenteel Onderzoek der
Materie'' (FOM). The work of R.L.~is partially supported by the Mexico's 
National Council for Science and Technology (CONACyT) under grant 010085. 
The work of N.A.-A.~and T.O.~is partially supported by the Spanish grant
FPA2000-1584.

\appendix
\section{Conventions}
\label{sec-conventions}

Greek indices $\mu,\nu,\rho,\ldots$ denote world tensor indices while 
Latin $a,b,c\ldots$ indices are tangent space indices.  We use hats for
11-dimensional objects and no hats for 8-dimensional objects. We
symmetrize and antisymmetrize with weight one.  Sometimes we use the
following convention: when indices are not shown explicitly, we assume
that all of them are world indices and all of them are completely
antisymmetrized in the obvious order. This is similar to
differential forms notation but the numerical factors differ.

We use mostly minus signature $(+-\cdots -)$\footnote{
  All formulae in this paragraph are valid both without and with hats.}. 
$\eta_{ab}$ is the Minkowski
spacetime metric and the spacetime metric is $g_{\mu \nu}$. Lorentz and world
indices are related by the Vielbeins $e_{a}{}^{\mu}$ and inverse
Vielbeins $e_{\mu}{}^{a}$, that satisfy
\begin{equation}
e_{a}{}^{\mu}e_{b}{}^{\nu}g_{\mu\nu}=\eta_{ab}\, ,
\hspace{1cm}
e_{\mu}{}^{a}e_{\nu}{}^{b}\eta_{ab}=g_{\mu\nu}\, .  
\end{equation}
The spin connection $\omega_{abc}$ is defined by
\begin{align}
\label{eq:spincon}
\omega_{abc} = -\Omega_{abc}+\Omega_{bca} -\Omega_{cab}\, , \qquad
\Omega_{ab}{}^{c} = e_{a}{}^{\mu}e_{b}{}^{\nu} \partial_{[\mu}e^{c}{}_{\nu]}\, .
\end{align}
The Riemann curvature tensor is given in terms of the spin connection by
\begin{align}
R_{\mu\nu a}{}^{b} = 2\partial_{[\mu}\, \omega_{\nu] a}{}^{b} 
-2\omega_{[\mu| a}{}^{c}\,\omega_{|\nu]c}{}^{b}\, . 
\end{align}  

The 11-dimensional gamma matrices satisfy the anticommutation relations
\begin{equation}
\{ \hat{\Gamma}{}^{\hat{a}},
\hat{\Gamma}{}^{\hat{b}}\} =
+2\hat{\eta}{}^{\hat{a}\hat{b}}\, ,
\end{equation}
We choose them to satisfy
\begin{equation}
  \hat{\Gamma}{}^{\hat{a}\, \star} = -\hat{\Gamma}{}^{\hat{a}} \,, \qquad
  \hat{\Gamma}{}^{0}\, \hat{\Gamma}{}^{\hat{a}}\,
  \hat{\Gamma}{}^{0}
  =\hat{\Gamma}{}^{\hat{a}\, \dagger}\, ,
\end{equation}
and thus are completely imaginary.
With the definition $\bar{\epsilon} = i \epsilon^\dagger \hat{\Gamma}{}^{0}$ 
we can derive the properties
\begin{equation}
\label{eq:spintrans}
\overline{\hat{\epsilon}}\,
\hat{\Gamma}{}^{\hat{a}_{1}\cdots \hat{a}_{n}}\,
\hat{\psi} =  
(-1)^{n+\left[n/2\right]}\, 
\bar{\hat{\psi}} \,
\hat{\Gamma}{}^{\hat{a}_{1}\cdots \hat{a}_{n}}\,
\hat{\epsilon}\, ,
\end{equation}
and so the above bilinear is symmetric for $n=0,3,4,7,8$ and antisymmetric for
$n=1,2,5,6,9,10$.

\section{$D=11$ Supergravity} \label{11Dsugra}

The bosonic part of the full $D=11$ supergravity reads in our conventions
\begin{align}
\hat{S} = & \frac{1}{16\pi G_{N}^{(11)}}{\displaystyle\int} d^{11}\hat{x}\
\sqrt{|\hat{g}|}\ \left[\hat{R}(\hat{\omega})
-{\textstyle\frac{1}{2\cdot 4!}}\hat{G}{}^{\, 2} 
-{\textstyle\frac{1}{(144)^{2}}} \frac{1}{\sqrt{|\hat{g}}|}
\hat{\epsilon}\hat{G} \hat{G} \hat{C}
\right]\, ,
\label{eq:11dtruncaction}
\end{align}

\noindent
where $\hat{G} = 4 \, \partial \hat{C}$ and $G_N^{(11)}$ is the
eleven-dimensional Newton constant.
This action is invariant under the local supersymmetry transformations
with parameter $\hat{\epsilon}$ 
\begin{align}
\delta_{\hat{\epsilon}} \hat{e}_{\hat{\mu}}{}^{\hat{a}}
= & 
-\frac{i}{2} \bar{\hat{\epsilon}}\, \hat{\Gamma}{}^{\hat{a}}\,
\hat{\psi}_{\hat{\mu}}\, , \notag \\
\delta_{\hat{\epsilon}} 
\hat{\psi}_{\hat{\mu}} = & 
 2\hat{\partial}_{\hat{\mu}} \hat{\epsilon}
 - \frac{1}{2} {\hat{\omega}}_{\hat{\mu}}{}^{\hat{a}\hat{b}}
\hat{\Gamma}_{\hat{a}\hat{b}} \hat{\epsilon}
+\frac{i}{144}
\left(
\hat{\Gamma}{}^{\hat{\alpha}\hat{\beta}
\hat{\gamma}\hat{\delta}}{}_{\hat{\mu}}
-8 \hat{\Gamma}{}^{\hat{\beta}
\hat{\gamma}\hat{\delta}}
\hat{\eta}_{\hat{\mu}}{}^{\hat{\alpha}}
\right)
\hat{\epsilon}\,
{\hat{G}}_{\hat{\alpha}\hat{\beta}
\hat{\gamma}\hat{\delta}} \, , \notag \\
\delta_{\hat{\epsilon}} 
\hat{C}_{\hat{\mu}\hat{\nu}\hat{\rho}}
= & 
\frac{3}{2} \bar{\hat{\epsilon}}\,
\hat{\Gamma}_{[\hat{\mu}\hat{\nu}}\,
\hat{\psi}_{\hat{\rho}]}\, ,
\label{eq:11Dsusy}
\end{align}
up to bilinears in fermions.

\section{The 8-Dimensional Bosonic Action} \label{full8daction}

Restricting to gauge groups with traceless 
structure constants, $f_{mn}{}^n = 0$, the full bosonic 8-dimensional action reads
\begin{align}
\label{eq:d8-Q-action}
S & = \frac{1}{16\pi G_{N}^{(11)}}\, C_U
{\displaystyle\int} d^{8}x \sqrt{|g_{E}|}\,
\left\{
R_{E} 
+\frac{1}{4}{\rm Tr}\left({\cal D} {\cal M}{\cal M}^{-1}\right)^{2}
+\frac{1}{4}{\rm Tr}\left(\partial {\cal W}{\cal W}^{-1}\right)^{2}
 \right. \notag \\
& 
\hspace{2cm}
-\frac{1}{4}F^{I\, m}{\cal M}_{mn}{\cal W}_{IJ} F^{J\, n}
+\frac{1}{2\cdot 3!} H_{m}{\cal M}^{mn} H_{n}
-\frac{1}{2\cdot 4!} e^{\varphi} G^{2} - {\cal V}\, \notag \\
& 
\hspace{2cm}
-\frac{1}{6^{3}\cdot 2^{4}}
{\textstyle\frac{1}{\sqrt{|g_{E}|}}}\, \epsilon
\left[GG \ell -8G H_{m} A^{2\, m} +12 G(\tilde{F}^{m}+\ell F^m)B_{m} \right. \notag \\
& 
\hspace{2cm}
\left.\left.
-8\epsilon^{mnp} H_{m}H_{n}B_{p} -8G\partial \ell C 
-16 H_{m}(\tilde{F}^{m}+\ell F^m) C \right]\right\} \, .
\end{align}
where we have made the following field strength definitions:
\begin{alignat}{2}
\label{eq:fieldstrengths2}
  G & = 4 \partial C + 6 F^{m} B_{m}\, , & \qquad
  F^{m} & = 2\partial A^{m} -\chi_{np}{}^{m}A^{n}A^{p}\, , \notag \\
  H_{m} & = 3{\cal D} B_{m} + 3 \epsilon_{mnp} F^{n} V^p\, , & 
  \tilde{F}^m & = 2{\cal D} V^m - \epsilon^{mnp} f_{np}{}^{q} B_{q}\, ,
\end{alignat}
The two vector field strengths form a doublet $F^{I\,m} = (F^m,\tilde{F}^m)$
with $I=1,2$. The $SL(2,\mathbb{R}) / SO(2)$ and $SL(3,\mathbb{R}) / SO(3)$
scalar cosets are defined in \eqref{Wscalar} and \eqref{Mscalar},
respectively. The potential ${\cal V}$ reads
\begin{equation}
 {\cal V} = {\textstyle\frac{1}{4}} e^{-\varphi}\,
             \left[ 2{\cal M}^{nq}f_{mn}{}^{p} f_{pq}{}^{m}
                  + {\cal M}^{mq}{\cal M}^{nr}{\cal M}_{ps}
                  f_{mn}{}^{p}f_{qr}{}^{s}
             \right]\, .
\label{eq:potential-1}
\end{equation}
The supersymmetric transformation rules in eight dimensions are
\begin{align}
\delta e_{\mu}{}^{a} = &
-\frac{i}{2}\overline{\epsilon} \Gamma^{a} {\psi}_{\mu} \, ,\notag \\
\delta \psi_\mu  = & 2 \partial_\mu \epsilon 
 - \frac{1}{2} {\slashed \omega}_\mu \epsilon
  +\frac{1}{2}L_{[i|}{}^{m} {\cal D}_{\mu} L_{m|j]}\Gamma^{ij} \epsilon \,
+\frac{1}{24}e^{-\varphi/2}f_{ijk}\Gamma^{ijk} \Gamma_\mu \epsilon \, \notag \\
& +\frac{1}{24} e^{\varphi/2}\Gamma^iL_{i}^{\ m}
( \Gamma_{\mu}^{\ \nu \rho}-10\delta_\mu^{\ \nu}\Gamma^\rho )
 F_{m\nu \rho}\epsilon +\frac{1}{2}e^{-\varphi}\partial_\mu \ell 
\epsilon \notag \\
& +\frac{i}{96}e^{\varphi/2}(\Gamma_{\mu}^{\ \nu \rho \delta \epsilon}
-4 \delta^{\ \nu}_{\mu} \Gamma^{\rho \delta \epsilon})
G_{\nu \rho \delta \epsilon} \epsilon 
+ \frac{i}{36}\Gamma^iL_i^{\ m}(\Gamma_{\mu}^{\ \nu \rho \delta}
-6 \delta^{\ \nu}_\mu \Gamma^{\rho \delta})H_{m \nu \rho \delta}\epsilon
\notag \\ 
& +\frac{i}{48}e^{-\varphi/2}\Gamma^i \Gamma^j L_i^{\ m} L_j^{\ n}
(\Gamma_\mu^{\ \nu \rho}-10\delta_\mu^{\ \nu}\Gamma^\rho)
(F_{mn \nu \rho} +\ell \varepsilon_{mnp}F^p_{\ \nu \rho})\epsilon \,, \notag \\
\delta \lambda_i  = &
 \frac{1}{2}L_i^{\ m} L^{jn}{\slashed {\cal D}}{\cal M}_{mn}
 \Gamma_j \epsilon -\frac{1}{3} {\slashed \partial} \varphi 
\Gamma_i \epsilon 
-\frac{1}{4}e^{-\varphi/2} (2f_{ijk}-f_{jki})\Gamma^{jk}\epsilon \, \notag \\
& + \frac{1}{4}e^{\varphi/2}L_i^{\ m}{\cal M}_{mn}{\slashed F}^n \epsilon +
\frac{i}{144}e^{\varphi/2}\Gamma_i {\slashed G}\epsilon 
+\frac{i}{36}(2\delta_i^{\ j}-\Gamma_{i}^{\ j})L_j^{\ m}{\slashed H}_m 
\epsilon \notag \\
& +\frac{i}{24}e^{-\varphi/2}\Gamma^j L_j^{\ m}L_k^{\ n}
(3\delta_i^{\ k}-\Gamma_{i}^{\ k})({\slashed F}_{mn}
+\ell \varepsilon_{mnp}{\slashed F}^p)\epsilon 
+\frac{1}{3}e^{-\varphi}\Gamma_i {\slashed \partial} \ell \epsilon \,,
\notag \displaybreak[2] \\
\delta A_{\mu}{}^{m}
= & -\frac{i}{2}e^{-\varphi/2} L_{i}^{\ m} \overline{\epsilon}  (
   \Gamma^{i} \psi_{\mu} -{\Gamma}_{\mu} 
(\eta^{ij}- \frac{1}{6}\Gamma^{i}\Gamma^{j})\lambda_j )  \,, \notag \\
\delta V_{\mu}^{\ m} = & -\frac{i}{2}e^{\varphi/2}
L_i^{\ m} \bar \epsilon ( \Gamma^i \psi_\mu
+\Gamma_\mu (\eta^{ij}-\frac{5}{6}\Gamma^i \Gamma^j) \lambda_j )-
\ell \delta A_\mu^{\ m}   \,, \notag \\
\delta B_{\mu \nu m} = & 
L_m^{\ \ i} \bar \epsilon  (\Gamma_{i[\mu} \psi_{\nu ]}
+\frac{1}{6} \Gamma_{\mu \nu}
(3\delta_i^{\ j}-\Gamma_i \Gamma^j)\lambda_j ) -2 \varepsilon_{mnp}
\delta A_{\mu}^{\ n} V_{\nu}^{\ p} \,, \notag \\
\delta C_{\mu \nu \rho} = & 
\frac{3}{2}e^{-\varphi/2} \bar \epsilon \Gamma_{[\mu \nu}( \psi_{\rho]}
-\frac{1}{6}\Gamma_{\rho]} \Gamma^i \lambda_i )
-3 \delta A_{[\mu}^{\ m} B_{ \nu \rho]m} \,, \notag \\ 
L^{\ n}_{i}\delta L_n{}_j = & \frac{i}{4}e^{\varphi/2}
  \overline\epsilon  (\Gamma_i \delta_j^{\ k} + \Gamma_j \delta^{\ k}_i-
  \frac{2}{3}\eta_{ij}\Gamma^k )\lambda_k \,,  \notag \\
\delta \varphi 
   = & -\frac{i}{2} \overline{\epsilon} \Gamma^{i} \lambda_i \,, \notag \\
\delta \ell = & -\frac{i}{2}e^{\varphi} \bar \epsilon \Gamma^i \lambda_i \,,
\end{align}

\section{The Bianchi  Classification}\label{classifications} \label{Bianchi}

In this Appendix we will discuss the Bianchi classification of 
three-dimensional Lie algebras. We will also show how different algebras are 
related via analytic continuation or group contraction. We compare our 
results with the $CSO(p,q,r)$ notation which is often used in the supergravity 
literature.

We assume that the generators of the three-dimensional Lie group
 satisfy the commutation relations ($m=1,2,3$)
\begin{align}
  [ T_m , T_n ] = f_{mn}{}^p T_p \,, 
\end{align}
with constant structure coefficients $f_{mn}{}^p$ subject to the Jacobi 
identity $ f_{m[n}{}^p f_{qr]p} = 0$. For three-dimensional Lie groups the 
structure constants have nine components and can be conveniently parameterized 
by
\begin{align}
  f_{mn}{}^p = \epsilon_{mnq} \mathsf{Q}^{pq} + 2 \delta_{[m}{}^p a_{n]} \,.
\end{align}
Here $\mathsf{Q}^{pq}$ is a symmetric matrix. The Jacobi identity implies 
$\mathsf{Q}^{pq} a_q =0$. Having $a_q=0$ corresponds to an algebra with 
traceless structure constants: $f_{mn}{}^n =0$.

Of course Lie algebras are only defined up to changes of basis $T_m 
\rightarrow R_m{}^n \, T_n$. This can always be used \cite{Wald:1984rg, 
Schirmer:1995dy} to transform $\mathsf{Q}^{pq}$ into a diagonal form and 
$a_q$ to have 
only one component. We will take $\mathsf{Q}^{pq} = \tfrac{1}{2} 
\text{diag}(q_1,q_2,q_3)$ and $a_q=(a,0,0)$. The commutation relations then 
take the form
\begin{equation}
   [T_1 , T_2] = \tfrac{1}{2} q_3 T_3 -a T_2 \,, \qquad
   [T_2 , T_3] = \tfrac{1}{2} q_1 T_1 \,, \qquad
   [T_3 , T_1] = \tfrac{1}{2} q_2 T_2 +a T_3 \,.
\end{equation}
The different three-dimensional Lie algebras have been classified and are 
given in Table \ref{3Dalgebras}. There are 11 different algebras, two of which 
are a one-parameter family. Of these only $SO(3)$ and $SO(2,1)$ are simple 
while the rest are all non-semi-simple \cite{Schirmer:1995dy,Hamermesh}. Note 
that for $a \neq 0$ the rank of $\mathsf{Q}$ can not exceed two due to the 
Jacobi identity.

\begin{table}[ht]
\begin{center}
\begin{tabular}{||c|c|c|c|c|c||}
\hline \rule[-3mm]{0mm}{8mm}
  Class & Bianchi & $a$ & $(q_1,q_2,q_3)$ & Group & $CSO(p,q,r)$  \\
\hline \hline \rule[-3mm]{0mm}{8mm}
  A & I & 0 & $(0,0,0)$ & $U(1)^3$ & $(0,0,3)$  \\
\hline \rule[-3mm]{0mm}{8mm}
  A & II & 0 & $(0,0,1)$ & Heisenberg & $(1,0,2)$  \\
\hline \rule[-3mm]{0mm}{8mm}
  A & VI$_0$ & 0 & $(0,-1,1)$ & $ISO(1,1)$ & $(1,1,1)$   \\
\hline \rule[-3mm]{0mm}{8mm}
  A & VII$_0$ & 0 & $(0,1,1)$ & $ISO(2)$ & $(2,0,1)$   \\
\hline \rule[-3mm]{0mm}{8mm}
  A & VIII & 0 & $(1,-1,1)$ & $SO(2,1)$ & $(2,1,0)$  \\
\hline \rule[-3mm]{0mm}{8mm}
  A & XI & 0 & $(1,1,1)$ & $SO(3)$ & $(3,0,0)$  \\
\hline \hline \rule[-3mm]{0mm}{8mm}
  B & V & 1 & $(0,0,0)$ & &  \\
\hline \rule[-3mm]{0mm}{8mm}
  B & IV & 1 & $(0,0,1)$ & &  \\
\hline \rule[-3mm]{0mm}{8mm}
  B & III & 1 & $(0,-1,1)$ & &  \\
\hline \rule[-3mm]{0mm}{8mm}
  B & VI$_a$ & $a$ & $(0,-1,1)$ & &  \\
\hline \rule[-3mm]{0mm}{8mm}
  B & VII$_a$ & $a$ & $(0,1,1)$ & &  \\
\hline
\end{tabular}
\caption{\label{3Dalgebras}\it The different three-dimensional Lie algebras in 
terms of components of their structure constants and the Bianchi and 
$CSO(p,q,r)$ classification. Note that there are two one-parameter families 
VI$_a$ and VII$_a$ with special case VI$_0$, VII$_0$ and VI$_1$=III. }
\end{center}
\end{table}

\begin{figure}[tb]
  \centerline{\epsfig{file=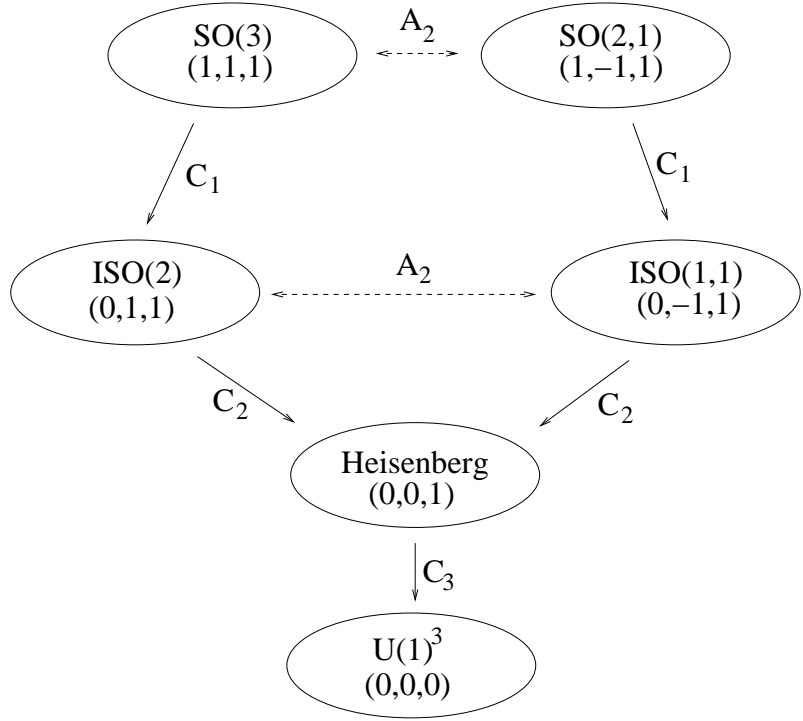,width=.65\textwidth}}
  \caption{\it Relations between groups associated to the 3D Class A
    Lie algebras. The boxes give the groups and the components
    $\mathsf{Q}^{mn} = \tfrac{1}{2} \text{diag}(q_1,q_2,q_3)$ of the
    structure constants. The arrows give the operations: the dashed
    arrow corresponds to the reversible analytic continuation, the
    solid arrow to the irreversible group contraction. These analytic
    continuations and contractions are defined in \eqref{C7} and
    \eqref{C8}.} \label{3Drelations}
\end{figure}

We will now show relations between all algebras of Class A, i.e. having $a=0$. 
Our starting point will be $SO(3)$. Its generators take, in our basis with 
$\mathsf{Q}=\tfrac{1}{2}\text{diag}(1,1,1)$, the form
\begin{align}
  T_1 = \tfrac{1}{2} \left( \begin{array}{ccc} 0 & 1 & 0 \\ 
    -1 & 0 & 0 \\ 0 & 0 & 0 \end{array} \right) \,, \qquad
  T_2 = \tfrac{1}{2} \left( \begin{array}{ccc} 0 & 0 & 1 \\ 
    0 & 0 & 0 \\ -1 & 0 & 0 \end{array} \right) \,, \qquad
  T_3 = \tfrac{1}{2} \left( \begin{array}{ccc} 0 & 0 & 0 \\ 
    0 & 0 & 1 \\ 0 & -1 & 0 \end{array} \right) \,.
\end{align}
One can obtain the other algebras with $a=0$ from these $SO(3)$ generators by 
the analytic continuation and/or contraction. Define the operations $A_1$ 
(analytic continuation) and $C_1$ (contraction) by
\begin{align}
  T_2 \rightarrow \lambda^{-1} \, T_2 \,, \qquad T_3 \rightarrow \lambda^{-1} \, T_3 \,, 
\end{align}
with $\lambda = i$ for $A_1$ and $\lambda \rightarrow 0$ for $C_1$. Its effect 
on the parameters of the algebra reads
\begin{align}
  \mathsf{Q} = \tfrac{1}{2}\text{diag} (q_1,q_2,q_3) \rightarrow 
\mathsf{Q} = \tfrac{1}{2}\text{diag} (\lambda^2 q_1,q_2,q_3) \,.
\end{align}
Thus from $SO(3)$ one can obtain $SO(2,1)$ by an $A$ operation and $ISO(2)$ 
by a $C$ operation. Similarly, the other Class A algebras are related by 
various analytic continuations and contractions, as shown in Figure 
\ref{3Drelations}.

It is instructive to compare the discussion of the previous paragraph
with the $CSO(p,q,r)$ notation which is often used in the supergravity
literature, see e.g.~\cite{Hull:1985rt, Hull:2002cv}. In our case
$p+q+r=3$ but the $CSO(p,q,r)$ classification of contracted algebras
is valid more generally. The $CSO(p,q,r)$ group is a group contraction
of $SO(p+r,q)$ and can be obtained as follows. One defines the
starting point $CSO(p,q,0) = SO(p,q)$. The effect of analytic
continuation in one of the $p$ directions is
\begin{align}\label{C7}
  A_p: \qquad CSO(p,q,r) \rightarrow CSO(p-1,q+1,r) \,,
\end{align}
while the effect of contraction is
\begin{align}\label{C8}
  C_p: \qquad CSO(p,q,r) \rightarrow CSO(p-1,q,r+1) \,.
\end{align}
This defines all Class A algebras in terms of the $CSO(p,q,r)$ classification, 
as shown in Table \ref{3Dalgebras}. These can all be obtained from the 
semi-simple algebras $SO(3)$ or $SO(2,1)$ by various contractions. Using the 
fact that $CSO(p,q,r) \sim CSO(q,p,r)$ one can see that Class A exhausts the 
possibilities of distributing $p,q,r$ subject to $p+q+r=3$.


\newpage
\bibliography{eight}
\bibliographystyle{toine}

\end{document}